# EXPLORING THE ASSOCIATION BETWEEN R&D EXPENDITURE AND THE JOB QUALITY IN THE EUROPEAN UNION

**Fernando Almeida**
University of Porto & INESC TEC
almd@fe.up.pt
**Nelson Amoedo**
Polytechnic Higher Institute of Gaya (ISPGaya)
nma@ispgaya.pt

## Abstract
*Investment in research and development is a key factor in increasing countries' competitiveness. However, its impact can potentially be broader and include other socially relevant elements like job quality. In effect, the quantity of generated jobs is an incomplete indicator since it does not allow to conclude on the quality of the job generated. In this sense, this paper intends to explore the relevance of R&D investments for the job quality in the European Union between 2009 and 2018. For this purpose, we investigate the effects of R&D expenditures made by the business sector, government, and higher education sector on three dimensions of job quality. Three research methods are employed, i.e. univariate linear analysis, multiple linear analysis, and cluster analysis. The findings only confirm the association between R&D expenditure and the number of hours worked, such that the European Union countries with the highest R&D expenses are those with the lowest average weekly working hours.*

## Introduction
In a knowledge-based economic context, the value of organizations does not only lie in their tangible assets (e.g. land, buildings, machinery), but also in immaterial or intangible aspects. These intangible elements include the ability to innovate, the ability to produce new products, the ability to establish partnerships, distribution channels, human capital, corporate culture, among other things (Niculita et al., 2012).
According to the OECD (2012), Research and Development (R&D) activities include systematic creative work to increase knowledge and use it in new applications. These activities can be subdivided into basic research, applied research, and experimental development. Basic research is defined as experimental or theoretical work undertaken with the primary aim of acquiring new knowledge about the foundations of phenomena or observable facts without any specific application in mind (OECD, 2012). In this sense, applied research differs from fundamental research in that it has a specific practical objective in mind. On the other hand, OECD (2012) characterizes experimental development as systematic work based on existing knowledge obtained through research or practical experience, to produce new or improved products, materials, or processes. Due to the distinctive characteristics of these three concepts, there are differences in how the results of the R&D process are disseminated. According to Cash and Culley (2015), the results of fundamental research are generally rapidly and widely disseminated, while the results of applied research and experimental development are disseminated under patent protection or licensing agreements.
Investments in R&D have been evidenced in the literature, over time, as one of the factors of technical progress and, therefore, contributing to the innovation and the increased productivity of countries (Blanco et al., 2013; Hornung, 2001). Furthermore, the R&D investment of one country also influences the productivity of other countries through spill-over effects and diffusion processes (Xing, 2018). In this sense, countries with strategies consolidated in policies that promote technological and non-technological development, stimulate innovation, education and advanced training are those that achieve higher levels of GDP per capita (Barkhordari et al., 2019; Diebolt and Hippe, 2019).
The European Union (EU) considers that investing in R&D is crucial for the future of Europe and enables the European social model to function (Kacprzyk and Doryn, 2017; Pece et al., 2015). The EU's support for R&D activities generates added value by encouraging collaboration between research teams from various countries and disciplines, which is central to the global scientific progress. In this sense, R&D activities in the EU increase the competitiveness of its member states, companies, and universities (Etzkowitz and Chou, 2017). Additionally, investments in R&D also contribute to improving the daily lives of millions of people in Europe and around the world (Crespi and Quatraro, 2015).
Simultaneously, one of the objectives of the Europe 2020 strategy is to create more and better jobs. Among the important principles, objectives and activities referred to in the Europe 2020 Strategy is the promotion of a high level of employment through the development of a coordinated strategy, in particular with a view to the creation of a skilled, qualified and adaptable workforce and labor markets capable of responding to economic change (EC, 2019). This vision is also shared by Piasna et al. (2019) who advocate the need to obtain indicators of job quality to be included in the EU's employment strategy. It is unequivocal that the quantified assessment of the volume of employment of nations needs to be supplemented by an analysis focusing on the quality of employment in all its dimensions. In this sense, this article aims to assess the relevance of R&D investments made by several entities (i.e., government, enterprises, and higher education institutions) for job quality in the EU, using data for the period 2009-2018. This article is innovative in exploring the issue of job quality as part of the Europe 2020 strategy and analyzing the relevance of its development from R&D investments. The aim is to find out whether these R&D investments have also helped to improve the living and working conditions of workers in the EU. Our main argument is that investment in R&D must be accompanied by the development of job quality, allowing simultaneously a European growth based on the simultaneous development of technological and social dimensions.
This paper is organized as follows: initially, a literature review of the role of R&D investment and the dimensions of job quality is carried out. The research questions are also defined in this section. Next, the methods used in this paper are presented. Subsequently, the results of the article are analyzed and discussed. Finally, the conclusions are discussed, and some indications of future work are given.

## Literature review

### Benefits and ways to stimulate R&D

Investments in R&D are a priority in several companies, and it is frequently observed that several companies leverage their competitive advantages through this type of investment. In fact, in the global and knowledge economy, the creation of innovation factors that allow companies to increase their level of competitiveness in the market is, in many cases, directly related to the availability that they have for the allocation of resources to R&D (Schmidt et al., 2016). According to Goffin and Mitchell (2016), companies invest in R&D for different reasons, such as the markets in which they operate, the size, the maturity of the business, the type of activity developed, the behavior of competition, or the market strategies outlined. The benefits of R&D activity can be felt more directly, through the products and processes resulting from innovation, or less immediately and more immaterial, through the development of skills and the ability to improve their performance (Dziallas and Blind, 2019).

The analysis of the benefits of R&D activity is explored by various actors. Lawson and Samson (2001) state the R&D strategy allows companies to create new innovative products or add ways to differentiate existing products from those offered by competitors. By adding new functionalities, due to a greater scientific and technological incorporation, it is expected that products and services can generate greater economic value. Furthermore, investment in R&D will also allow the development of new, more efficient production methods that reduce the cost of production or provision of services (Özer, 2012). In this sense, investment in R&D may result in more efficient production processes, which reduce marginal production costs or allow the creation of more efficient products with higher marginal productivity. Anadon et al. (2016) advocate this greater efficiency globally promotes the decoupling of economic growth from the increase in resource consumption, promoting sustainability.

It is also important to note that investment in R&D may not have an immediate practical effect. R&D effort is necessarily an activity with a high level of uncertainty, in which the return on investment is assumed to be difficult to predict (Spacek and Vacik, 2016). However, the accumulated knowledge is an asset of the company which, if well managed, can become a competitive advantage. Podmetina et al. (2018) refer that the scientific and technological knowledge generated by R&D activities allow companies to develop new skills and increase their agility to anticipate trends and be able to change their strategy. Additionally, several companies are operating in the industrial property market that allows them to trade patents and royalties through the developed R&D activities (Attorney, 2017).

The European Union and each Member State are interested in and committed to fostering investment in business R&D through the definition of concrete objectives and policies, on the understanding that this will eventually translate into tangible gains in competitiveness. There are multiple forms of public support for corporate R&D and all EU countries have public policies to stimulate companies to invest in R&D, although with relevant differences considering the specificities of each country and the mix of instruments that materialize such support. In the European context, two major groups of public support instruments stand out: direct funding and indirect funding (Van Pottelsberghe et al., 2003). Direct funding of R&D projects is provided through grants, loans, or other financial instruments. On another perspective, indirect financing is carried out through tax incentives that reduce the amount of tax to be paid by companies. According to Westmore (2014), direct financing and tax incentives are instruments with different characteristics and are only partially substitutable with each other. Traditionally, public policies have favored direct financing over corporate R&E. However, recent data provided by the OECD (2018) point to a growing increase in the number of countries with tax incentive systems for R&D, with a progressive replacement of direct funding by tax incentives.

Another way to encourage R&D activities is through public/private higher education institutions. Universities are reference points in building the design of science and are also promoters of innovation that contribute to the economic development of countries. The triple helix model proposed by Etzkowitz and Leydesdorff (2000) argues that universities can play an increasingly relevant role in innovation in the context of knowledge-based societies. Briefly, the triple helix model expresses a new configuration of emerging institutional forces in support systems for R&D activities. Hunady et al. (2019) state that in a knowledge-based economy, the university becomes central in the process of providing human capital and in the process of launching new enterprises sustained in the growth generated within universities and research centers. In this sense, several countries have adopted measures to strengthen the relationship between the university and industry, to channel the contributions of academic research to innovation and fostering the economy. These measures focus on the strategic actors associated with the process of knowledge transfer and on the permissiveness that universities can support innovation in the industry through the production of products and/or processes that can be marketed (Philippi et al., 2018).

## Perspectives about the job quality

The new information economy has created paradoxical challenges for the European Union countries. From one perspective, some European citizens are unemployed and cannot find a job, and on the other hand, some citizens need to work more than 60 hours a week through various work activities. According to Gautier and Moraga-González (2018), the number of hours of work and the intensity of workload is currently higher than in the past. This vision is complemented by Imran et al. (2015) when highlighting that job security, loyalty and commitment of employees is no longer an acquired factor, as it was in organizations of the past. Therefore, companies must empower and motivate their employees to increase their competitiveness.

Another facet of the literature has emerged that aims to complement the traditional view of the number of new jobs created. This perspective seeks to assess the level of job quality, thus complementing the analysis of the volume of employment with another that focuses on the quality of these jobs. Job quality emerges as a necessarily multidisciplinary concept in which various perspectives on citizens' employment should be analyzed together. In the literature, it is possible to find several proposals for job quality indices that contemplate objective and subjective dimensions, in which the physical conditions of work, the health of employees, learning, the quality of the employer-employee relationship, etc. are found. (Amossé and Kalugina, 2010; Tangian, 2005). Additionally, Leontaridi and Sloane (2001) propose the existence of subjective indicators obtained from the use of job satisfaction surveys.

One of the essential aspects of job quality is the length of working hours. Traditionally, the number of working hours is associated with the success, commitment, and productivity of a job. However, this view is contradicted by Carmichael (2015) who compiled several studies in the area and concluded that spending more hours at work can be harmful to both the worker and the company. Additionally, this study states that managers could not distinguish the most productive workers from the least productive by looking exclusively at working hours. Statistics released by the OECD indicate that countries with longer working hours do not necessarily have higher productivity (OECD, 2019). On the contrary, studies conducted by Lepinteur (2016) and Glaveski (2018) indicate that shorter working hours favor productivity and contribute to reconciling employment and family. In this sense, it is important to explore whether the investment in R&D has led to a reduction in the number of working time hours, promoting a work-life balance. Therefore, the following hypothesis was formulated:

*H0: In EU countries, higher levels of R&D expenses are negatively associated with the average weekly working hours*

The labor market in recent years has undergone various transformations, from the process of technological innovation, changes in production processes, changes in organizational structures, as well as in human resources management. Fudge and Strauss (2013) argue that in a global economy, with the predominance of financial logic and short-term profitability, there is continuous pressure for labor factor flexibility. In response to these challenges and the seasonality of demand in certain sectors of activity (e.g., tourism), companies have been changing their operations, namely using new forms of hiring workers, increasingly resorting to temporary work. Temporary work is a job with a defined temporary limitation, with some asymmetries in the EU countries, although the average reference value is up to a maximum of two years (Eichhorst et al., 2017). However, the definition of temporary work is more complex and is based on a tripartite or triangular relationship between the worker, the temporary employment agency, and the company where the worker is currently going to work during the period of this relationship (Schoukens and Barrio, 2017). In this sense, and given the increasing importance of temporary work in recent years in the EU, it is important to explore whether the investment in R&D has contributed to increasing employment stability. Therefore, the following research question was established:

*H1: In EU countries, higher levels of R&D expenses are negatively associated with the percentage of temporary work*

Another factor impacting the quality of work is the precariousness of labor relationships. Hipp (2016) includes in this component a set of different labor configurations, such as service contracts, seasonal work without contracts, and occasional work. Campbell and Price (2016) state that considering the historical dimension of this problem, job insecurity has particularly affected the working classes. However, the profile of the precarious worker has been broadened to include young people under 25, women, and less qualified professional groups (Pembroke, 2018). Furthermore, job insecurity is more prevalent in seasonal economic sectors and small businesses (ILO, 2015). In fact, we are facing a situation where more and more individuals are at risk of social and economic precariousness. The study carried out by Oliveira and Carvalho (2008) allows us to conclude that precarious work affects EU countries differently and has been a structural feature in the reconfiguration of labor markets. In this sense, and given the importance that job insecurity has on the quality of life of citizens in the EU, it is important to explore whether the investment in R&D has contributed to reducing the rate of job insecurity. Therefore, the following research question was established:

*H2: In EU countries, higher levels of R&D expenses are negatively associated with the percentage of precarious work*

With the enlargement of Europe to 27 member states in January 2007, the EU increased its geographical coverage and population, but inevitably the asymmetries between its member states emerged. The economic, social, demographic, and territorial disparities in the EU-27 are widening, and the gap between the center and the periphery, between the richest and poorest regions, has increased (Götz; 2016; Scharpf, 2009). Furthermore, the global financial crisis of 2007-2012, which affected the EU mainly in the period 2009-2013, was one of the factors that increased the asymmetries, especially between the countries of southern and northern Europe. In this sense, it is pertinent to explore the asymmetries of each of the EU-27 countries, considering the three indicators of job quality previously considered. Therefore, the following research question was established:

*H3: In EU countries, higher levels of R&D expenses are associated positively with the levels of job quality*

# Methodology

## Data

The data for this study come from Eurostat, which is the official US statistical office. The data available at Eurostat come from the various Member States through the European Statistical System (ESS) and allow comparison of statistical data from different EU countries. Croatia was not included in this study because joined the European Union on 1st July 2013.

Firstly, the research and development expenditure database was accessed. According to Eurostat (2019), R&D expenditure is defined as: “Research and experimental development (R&D) comprise creative work undertaken on a systematic basis to increase the stock of knowledge, including knowledge of man, culture and society, and the use of this stock of knowledge to devise new applications. R&D expenditures include all expenditures for R&D performed within the business enterprise sector (BERD) on the national territory during a given period, regardless of the source of funds. R&D expenditure in BERD is shown as a percentage of GDP (R&D intensity).” Furthermore, to have a greater granularity of the analysis, three perspectives were considered: (i) business enterprise sector; (ii) government sector; and (iii) higher education sector. This information was complemented with the total number of researchers for the three sectors of activity identified above. Eurostat (2019) defines total researchers by sector of performance as “Researchers are professionals engaged in the conception or creation of new knowledge, products, processes, methods and systems, and in the management of the projects concerned. FTE (Full-time equivalent) corresponds to one year's work by one person (for example, a person who devotes 40 % of his time to R&D is counted as 0.4 FTE.” The role of R&D expenditure considering its multiple perspectives has been covered in several studies seeking to explore the competitiveness of EU countries (Kiselakova et al., 2018; Pelikánová, 2019). These three data views correspond to six independent variables, respectively:

- RDB – R&D expenditure (% GDP) in the business enterprise sector;
- RDG – R&D expenditure (% GDP) in the government sector;
- RDHE – R&D expenditure (% GDP) in the higher education sector;
- RSB – % of researchers in the business enterprise sector;
- RSG – % of researchers in the government sector;
- RSHE – % of researchers in the higher education sector.

Figure 1 visually presents the evolution of the R&D expenditure share of GDP and the percentage of researchers in the population considering the three data views defined above. The behavior of these variables allows us to conclude that R&D expenditure and the number of researchers in the business enterprise sector has gradually increased over the last ten years, while the contribution of the government sector and higher education sector has stagnated, remaining relatively constant over the period, with small fluctuations mainly in R&D expenditure in the higher education sector. Furthermore, it should be noted that the business enterprise sector has been the main contributor to R&D activities in the EU. This vision is confirmed by Berzkalne and Zelgalve (2012) and Hana (2013) that establish the importance of R&D investments in the competitiveness of companies through the creation of new products, services, or processes.

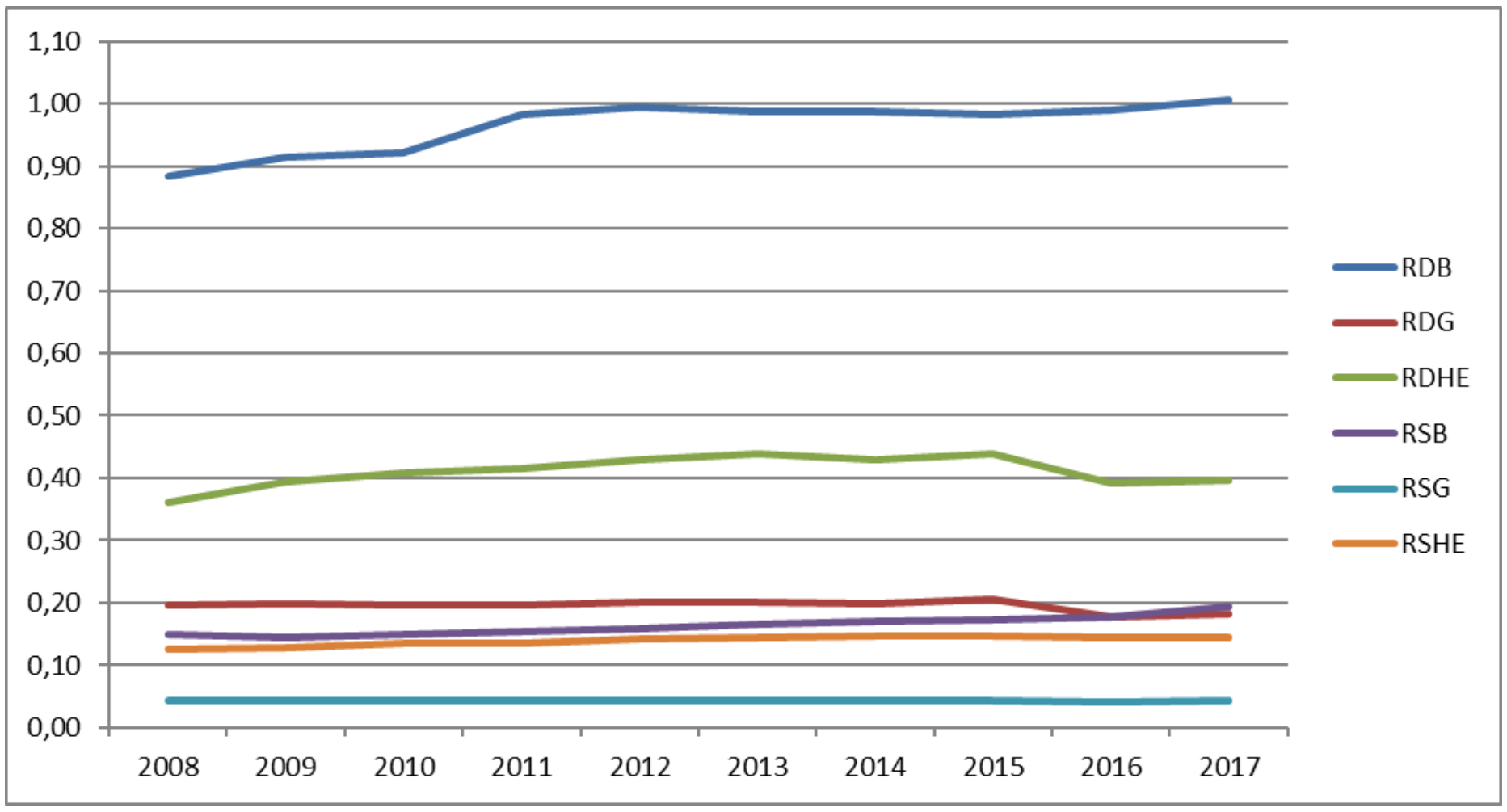

**Figure 1 Evolution of R&D dimensions in the EU**
Source: Eurostat (2019). *Eurostat – Your key to European statistics*. Available at: https://ec.europa.eu/eurostat/ (accessed 15th July 2019).

Secondly, it is also necessary to obtain data on the quality of work by considering the following three dimensions:

- Average number of usual weekly hours of work in the main job (WH). The WB has been explored in studies that intend to investigate the impact of long working hours on job quality (Artazcoz et al., 2018). The average number of working hours in the EU has remained relatively steady despite a trend over the last 10 years of a small reduction (average for 2009 was 38.3 hours and for 2018 is 37.9 hours). There are relevant fluctuations with Greece as the country with the highest number of 42-hour working hours for 2018 and the Netherlands with a value of 30.4 hours;
- Temporary employees as a percentage of the total number of employees (TE). The role of TE has been studied from the perspective of exploring its relevance in the economic recovery of the EU countries (ter Weel, 2018). According to Eurostat (2019), a person is considered a temporary worker if he/she is working for a temporary work agency. Temporary work has been growing progressively at an average annual rate of 6.9% since 2009. Its maximum values were reached in 2015 and 2016 with a percentage of 12.1 and 12.0, respectively. However, since this period the percentage of temporary workers has been decreasing, and in 2018 the percentage of workers in this situation is 11.6%.
- Precarious employment as a percentage of the total number of employees (PE). The role of PE has been studied from the perspective of its impacts on lifestyle and health conditions in the EU (Matilla-Santander et al., 2020). Eurostat (2019) establishes that a precarious worker has a short-term contract of up to 3 months. The number of workers in this situation has experienced some increase and represents approximately 2% of workers in the EU. The effects of the global financial crisis in the EU had some impact on this indicator in the period 2010-2015, where there was a growth rate higher than 22% compared to 2009.

Table 1 shows the evolution of the dependent variables (i.e., WH, TE, PE) in the period from 2009 to 2018. Table 1 omits the specificities of each country and presents only the EU-27 average for each year (Croatia was excluded). Of the three indicators, ET is the one with the highest asymmetry between countries. For 2018, the number of temporary agency workers in Spain was 26.9%, while the number of temporary agency workers in Romania was only 1.1%.

**Table 1 Evolution of WH, TE, and PE from 2009 to 2018**

| Year | WH | TE | PE |
|---|---|---|---|
| 2009 | 38.3 | 10.8 | 1.8 |
| 2010 | 38.2 | 11.4 | 2.0 |
| 2011 | 38.1 | 11.6 | 2.2 |
| 2012 | 38.0 | 11.3 | 2.1 |
| 2013 | 37.9 | 11.5 | 2.0 |
| 2014 | 37.9 | 11.9 | 2.1 |
| 2015 | 37.9 | 12.1 | 2.2 |
| 2016 | 37.9 | 12.0 | 2.1 |
| 2017 | 37.9 | 11.9 | 2.1 |
| 2018 | 37.9 | 11.6 | 2.0 |

Source: own source

## Methods

This article uses a quantitative approach for data exploration and analysis. Quantitative research allows exploring the relationships between the variables by experimental or semi-experimental methods through the adoption of various statistical techniques (Swift and Piff, 2014). Furthermore, Queirós et al. (2017) advocate this methodology also allows confirming the research hypotheses formulated by the researcher and assesses the degree of robustness and reliability of the obtained results. The SPSS v.21 was used to explore the relationship between the data. The following quantitative methods were applied:

- Univariate linear regression – a statistical methodology that uses the relationship between two quantitative variables to predict the behavior of a predictive variable from an independent variable (Montgomery et al., 2012). The simplest case of linear regression is when we have two variables and the relationship between them can be represented by a straight line. As suggested by Montgomery et al. (2012) the outliers were removed to avoid negatively biasing the result of the linear estimation;
- Multiple linear regression – an extension of the simple linear regression in which several predictors are used to estimating the value of the dependent variable (Montgomery et al., 2012). However, in multiple regression models, it is very common the occurrence of highly correlated independent variables which originates estimated regression coefficients with low precision. In these cases, it is useful to reduce the dimensionality of the model. One of the approaches is the adoption of principal component analysis (Jolliffe and Cadima, 2016);
- Cluster analysis – a multivariate statistical procedure that serves to identify homogeneous groups in data, based on variables or cases. Cluster analysis allows classifying objects based on the observation of similarities and dissimilarities (Everitt et al., 2011). According to King (2014), the cluster analysis should include the following phases: (i) selection of the elements to be grouped; (ii) definition of a set of variables from which the necessary information to group the cases will be obtained; (iii) selection of a measure of similarity or distance between each pair of cases; (iv) choice of an aggregation criterion; and (v) validation of findings.

## Results

### Simple linear analysis

Table 2 presents a simple linear correlation analysis between the variables under study. An analysis of variance (ANOVA) was performed to compare the distribution of several groups in independent samples. The F test is used to test the hypothesis that the regression model fits the proposed linear models (Warne, 2017). A significance level of 5% ($\alpha$=0.05) was adopted. The results suggest a good fit between the investments in R&D performed by companies and the higher education sector and the number of working hours. These cells are highlighted in Table 2. The correlation coefficient for business R&D investment is higher than 0.7, which indicates a strong correlation between the two variables, while the correlation for R&D expenditure by the higher education sector is moderate ($0.40 \leq R \leq 0.69$). However, it was not possible to establish the same association for TE and PE. The Durbin-Watson test was also conducted to test for the presence of autocorrelation in the errors of a regression model (Warne, 2017). It is referred by Warne (2017) that a Durbin-Watson score around 2 indicates non-autocorrelation; a value near 0 indicates positive autocorrelation among the variables; and a negative autocorrelation can be found in a value near 4. In situations where there is a good fit in the model, we can verify that the Durbin-Watson statistic is higher than 2.035 (tabulated reference value considering the sample size and number of terms) indicating there is no autocorrelation between the errors of the linear estimate.

**Table 2 Simple linear analysis**

| Variable | R | R Square | t-value | Durbin-Watson |
|---|---|---|---|---|
| | | *WH (df1 = 1 & df2 = 24)* | | |
| RDB | 0.702 | 0.492 | 0 | 2.088 |
| RDG | 0.086 | 0.007 | 0.676 | 1.899 |
| RDHE | 0.680 | 0.463 | 0 | 2.240 |
| RSB | 0.745 | 0.556 | 0 | 2.167 |
| RSG | 0.194 | 0.038 | 0.342 | 1.911 |
| RSHE | 0.500 | 0.250 | 0.009 | 2.258 |
| | | *TE (df1 = 1 & df2 = 24)* | | |
| RDB | 0.144 | 0.021 | 0.482 | 1.749 |
| RDG | 0.040 | 0.002 | 0.847 | 1.675 |
| RDHE | 0.101 | 0.010 | 0.623 | 1.712 |
| RSB | 0.074 | 0.005 | 0.721 | 1.704 |
| RSG | 0.170 | 0.029 | 0.406 | 1.775 |
| RSHE | 0.368 | 0.135 | 0.064 | 1.712 |
| | | *PE (df1 = 1 & df2 = 25)* | | |
| RDB | 0.327 | 0.107 | 0.096 | 1.711 |
| RDG | 0.266 | 0.071 | 0.181 | 1.788 |
| RDHE | 0.181 | 0.033 | 0.365 | 1.649 |
| RSB | 0.259 | 0.067 | 0.192 | 1.659 |
| RSG | 0.177 | 0.031 | 0.378 | 1.668 |
| RSHE | 0.110 | 0.012 | 0.583 | 1.669 |

Source: own source

## Multiple linear analysis

Multiple regression is an appropriate statistical method of analysis when the behavior of a dependent variable is related to several independent variables. According to Montgomery et al. (2012), the objective of the regression analysis is to predict the changes in the dependent variable in response to the changes that occur in the various independent variables. In this study, six independent variables (i.e., RDB, RDG, RDHE, RSB, RSG, RSHE) and three dependent variables (i.e., WH, TE, PE) are considered and tested separately. The principal component analysis was adopted to reduce the number of initial variables without significant loss of information (Jolliffe and Cadima, 2016). The authors sought to organize the data into three main components (i.e., CP1: RDB + RSB, CP2: RDG + RSG, CP3: RDHE + RSHE) to explore the implicit relationship between R&D expenditure and the number of researchers. However, this initial approach proved to be inadequate because the Variance Inflation Factor (VIF) was higher than 10. According to Darlington and Hayes (2016), this situation indicates a high correlation between the components. Therefore, the components were reduced to two (i.e., CP1: RDB + RDHE + RSB + RSHE, CP2: RDG + RSG), which obtained a VIF equal to 3.543. This value is acceptable for multiple linear regression purposes, revealing a relatively low correlation between the components (Darlington and Hayes, 2016).

Table 3 shows the results of multiple linear regression considering CP1 and CP2 for the three dependent variables (i.e., WH, TE, and PE). The analysis was carried out considering the period 2009-2018, because as stated by Aghion and Jaravel (2015) and Pradeep et al. (2017) the effects of an investment in R&D have economic effects in the medium and long term, so it is expected that the results of this investment will not be visible in the year in which this investment is made. The adjusted R squared is calculated because it provides a vision of the goodness-of-fit for regression models that contain different numbers of independent variables. The findings indicate a good fit of the model (P-value $< 0.05$) only to estimate the behavior of the dependent variable WH. The adjusted R2 is only relevant for WH, which indicates a moderate correlation between the independent components and the dependent variable WH. The association between R&D, TE, and PE was not supported. Finally, it is also pertinent to analyze the behavior of the degrees of freedom (df) since the outliers were removed. The removal of outliers occurred mainly in the regression analysis of TE and PE due to the asymmetric behavior of several European Union countries (e.g. Spain has a PE percentage approximately 25 times higher than Lithuania or Slovenia). In the estimation of PE and TE, particularly in the latter case, the number of data to estimate the parameter values was much smaller due to the high presence of outliers.

**Table 3 Multiple linear analysis**

| | Dependent variable | | |
|---|---|---|---|
| | WH (df1 = 2 & df2 = 20) | TE (df1 = 2 & df2 = 12) | PE (df1 = 2 & df2 = 15) |
| CP1 | 0.279 | 0.059 | 0.016 |
| CP2 | 0.403 | 0.167 | 0.068 |
| Mean | 38 | 11.61 | 2.06 |
| Std. dev. | 0.149 | 0.390 | 0.117 |
| $R^2$ | 0.480 | 0.409 | 0.254 |
| Adjusted $R^2$ | 0.426 | 0.257 | 0.100 |
| F | 8.865 | 2.846 | 1.578 |
| P-value | 0.002 | 0.151 | 0.392 |
| Durbin-Watson | 1.765 | 1.679 | 1.801 |

*P < 0.05

Source: own source

## Cluster analysis

Cluster analysis is a statistical technique used to classify elements and organize them into groups. The idea is that elements within the same cluster are very similar, and elements in different clusters are distinct from each other. In the definition of similarity and difference between countries, the Euclidean distance was used, in which the distance between two cases is the square root of the sum of the squares of the differences between the cases for all the variables (Mortier et al., 2006). Cluster analysis is pertinent due to the asymmetrical behavior of European Union countries. In this sense, this cluster analysis aims to understand specific behaviors and explore whether the association between the variables under study is relevant within each cluster. In Table 4 the countries were grouped into a total of 6 clusters according to the three dependent variables (i.e., WH, TE, PE).

**Table 4 Correspondence between countries and clusters**

| Countries | Clusters | | | | | |
|---|---|---|---|---|---|---|
| | **WH** | | **TE** | | **PE** | |
| | A | B | C | D | E | F |
| Austria | | X | X | | X | |
| Belgium | | X | X | | | X |
| Bulgaria | | X | X | | X | |
| Cyprus | | X | X | | X | |
| Czechia | | X | X | | X | |
| Denmark | X | | X | | X | |
| Estonia | | X | X | | X | |
| Finland | | X | | X | | X |
| France | | X | | X | | X |
| Germany | X | | X | | X | |
| Greece | | X | X | | X | |
| Hungary | | X | X | | X | |
| Ireland | | X | X | | X | |
| Italy | | X | | X | | X |
| Latvia | | X | X | | X | |
| Lithuania | | X | X | | X | |
| Luxembourg | | X | X | | X | |
| Malta | | X | X | | X | |
| Netherlands | X | | | X | X | |
| Poland | | X | | X | | X |
| Portugal | | X | | X | | X |
| Romania | | X | X | | X | |
| Slovakia | | X | X | | X | |
| Slovenia | | X | | X | | X |
| Spain | | X | | X | | X |
| Sweden | | X | | X | | X |
| United Kingdom | | X | X | | X | |

Source: own source

The size of each cluster is relatively identical, except for the WH. Here it appears that cluster 1 has only 3 countries (i.e. Denmark, Germany, and Netherlands) that stand out because they have hours of work far below the average of the other EU countries. Cluster D and F contain the countries with the highest percentage of temporary and precarious work. After the organization of the cluster countries, the quality of the adjustment and the degree of relationship between the independent and dependent variables in the study were tested for each year (Table 5). There are significant differences in behavior between clusters due to the heterogeneity of the EU countries and emerge some slight fluctuations in the quality of adjustment and in the linear correlation over the years within the same cluster. The level of significance of the quality of the adjustment is less than 0.05 in cluster A for the dependent variable WH throughout the considered period (i.e., 2009-2018). This indicates there is an association between R&D and WH in countries where the number of working hours is lower, i.e. R&D investments have contributed to reducing the number of working hours. However, the same relationship is not established for cluster B countries. There is only one more point situation (i.e., cluster C in the year 2015) where Sig. F. Change is less than 0.05 but does not allow drawing conclusions for the entire 2009-2018 period. It is pertinent to highlight that Durbin-Watson values are less than 1 in some clusters (e.g., B and F) in specific years. However, and considering the whole period of 10 years, its impact is residual.

**Table 5 Clusters analysis**

| Variable | Cluster | R Square | Adjusted R Square | F Change | Sig. F Change | Durbin-Watson |
|---|---|---|---|---|---|---|
| WH | A | 0.521 | 0.435 | 8.298 | 0.007 | 1.786 |
| | B | 0.163 | -0.306 | 0.423 | 0.518 | 1.667 |
| TE | C | 0.313 | 0.234 | 4.114 | 0.087 | 1.750 |
| | D | 0.327 | 0.194 | 2.695 | 0.112 | 1.812 |
| PE | E | 0.048 | -0.014 | 0.801 | 0.406 | 1.895 |
| | F | 0.191 | -0.052 | 0.958 | 0.383 | 1.497 |

Source: own source

## Discussion

The findings revealed a relationship between R&D expenditure and the number of working hours. In this sense, it was possible to confirm the H0 hypothesis that the EU countries with the highest R&D expenses are also those with lower working hours. However, this relationship only applies to R&D expenditures by the business enterprise sector and the higher education sector. The results did not allow us to confirm this relationship for the R&D expenditures carried out by the government sector. According to Walterskirchen (2016), a country's productivity increase can be achieved by reducing working hours or increasing real wages. Furthermore, Pullinger (2014) highlights the role of reducing the number of working hours in the economic development and sustainability.

From another perspective, it was not possible to confirm the H1 hypothesis, which means it was not feasible to establish a link between R&D expenses and the percentage of temporary work. The role of temporary work in the economy is not consensual. From one perspective, Vergeer and Kleinknecht (2014) state the importance of temporary work in reducing labor market rigidities and increasing productivity growth, while the findings to the study by Lisi and Malo (2017) mention that temporary work has a negative impact on productivity growth, particularly in skill intensity fields. Therefore, the relevance of temporary work assumes a very heterogeneous role for each EU country, with some countries, such as Romania, Bulgaria and Baltic countries, where the average TE in the last 10 years was less than 5%, while other countries, such as Spain, Poland, and Portugal, exceed 20%. These asymmetries are very relevant and explain the contradictory role of the TE in the EU. In the same direction, it was also not possible to confirm the H2 hypothesis, i.e. it was not feasible to establish a relationship between R&D expenses and precarious work. Higher investments in R&D have not contributed to a reduction of precarious work. The fight against precarious work is highlighted in European policy as a fundamental element in promoting the quality of employment and in combating precarious lives and health problems (Matilla-Santander, 2020; Pembroke, 2019), but the investments in R&D have not made a decisive contribution for this goal.

Finally, cluster analysis contributed to identifying some patterns of behavior among EU countries. It was possible to identify six clusters considering the number of working hours, and the percentage of temporary and precarious workers. This heterogeneity of behavior in European Union countries was also previously studied by Weresa (2019) in her analysis of the impact of the fourth industrial revolution. In this sense, the data from this study also confirm the heterogeneity and relevance of cluster analysis. The relationship between R&D expenses and the number of working hours is mainly relevant for countries with fewer working hours, i.e. those in cluster 1 (i.e. Denmark, Germany, and the Netherlands) with less than 35 working hours per week. In cluster A, the Netherlands stands out with an average working time of 30.4 hours over the last 10 years. However, this does not mean that workers in the Netherlands are in precarious employment, as this rate is one of the lowest in the EU. Dodds (2017) points out that this statistic is a result of the work culture in the Netherlands, whose employees like to set aside a good amount of time for family life and parallel activities. However, this relationship could not be confirmed for cluster B, indicating that this relationship cannot be confirmed for countries with small fluctuations in working hours. Finally, the cluster study carried out for the TE and PE components did not contribute to confirm these relationships independently of the considered clusters. Thus, the H3 hypothesis could not be confirmed for the three considered dimensions (i.e., WH, TE, and PE).

## Conclusions

The labor market in the EU has undergone major changes, in which companies tend to be increasingly challenging. Through this, employees seek to meet the standards and expectations of organizations, always seeking to meet the challenges of the market and live constantly under pressure. For organizations, it is essential to have motivated and committed human resources, therefore, quality of life in organizations becomes an indispensable factor to make employees feel motivated and committed to the objectives and goals of the organization, bringing benefits for the growth and success of the company.

The job quality has become the main part of competitiveness in the market, contributing to increasing the employees' performance and productivity. This study sought to explore the role of R&D expenditure in the work quality in the European Union in the period 2009-2018. The results identified a relationship between the increase in R&D expenditure and the number of worked hours, and this association is particularly relevant for countries with the lowest number of working hours in the EU (i.e. Denmark, Germany, and the Netherlands). However, it was not possible to confirm this relationship for temporary and precarious work.

This work has both theoretical and practical implications. From a theoretical point of view, it establishes the association between R&D expenditure, weekly working hours, the percentage of temporary work, and the percentage of precarious work. It also explores the relevance of the role of R&D expenditure by companies, government, and the higher education sector. From a practical point of view, the findings are relevant for the establishment of support policies that promote the improvement of the job quality in the EU. However, there are some limitations of this article that should be highlighted, namely: (i) the article considers only the value of R&D expenditure not looking at the quality and impact of that investment on the market; (ii) the value of precarious work was taken from official statistics provided by Eurostat, ignoring the existence of informal and undeclared precarious work; and (iii) cluster analysis allows only a superficial analysis of the impact of each country's asymmetry in the EU.

As future work, the authors intend to explore other elements in the quality of work, such as non-wage benefits, social protection, working conditions, skills development and training, or work motivation. These elements are relevant to be considered as control variables in the multiple regression model. Furthermore, we intend to explore regional aspects of the behavior of some countries to have a more detailed and deeper analysis of clusters, which may consider the regional and contextual specificities of each country. In this sense, it is suggested that the analysis could be more focused on the light of the asymmetries identified in the European Union countries.